\documentclass[conference]{aircc}

\usepackage{cite}
\usepackage{amsmath,amssymb,amsfonts}
\usepackage{algorithmic}
\usepackage{graphicx}
\usepackage{textcomp}
\usepackage{xcolor}
\usepackage{multirow}
\usepackage[hyphens]{url}
\usepackage{wrapfig}
\usepackage{graphicx}
\usepackage{lipsum}
\input{insbox}
\usepackage[export]{adjustbox}
  
\begin{document}

\title{Investigating Multi-Feature Selection and Ensembling for Audio Classification}
\author{Muhammad~Turab\textsuperscript{1}, Teerath~Kumar\textsuperscript{2,3}, Malika~Bendechache\textsuperscript{2,3,4}, and Takfarinas~Saber\textsuperscript{4,5}
}
\affiliation{
    \textsuperscript{1} Mehran University of Engineering and Technology, Jamshoro, Pakistan.
    \email{turabbajeer202@gmail.com} \hspace{10cm}
    \textsuperscript{2}~ADAPT~--~Science~Foundation Ireland~Research~Centre \hspace{10cm}
    \textsuperscript{3}~CRT~AI,~School of Computing, Dublin City University, Dublin, Ireland
    \email{teerath.menghwar2@mail.dcu.ie, malika.bendechache@dcu.ie} \hspace{10cm}
    \textsuperscript{4} Lero – the Irish Software Research Centre \hspace{10cm}
    \textsuperscript{5} School of Computer Science, National University of Ireland, Galway, Ireland \hspace{10cm} 
    \email{takfarinas.saber@nuigalway.ie} \hspace{10cm}
    }

\maketitle
\begin{abstract}
Deep Learning (DL) algorithms have shown impressive performance in diverse domains. Among them, audio has attracted many researchers over the last couple of decades due to some interesting patterns--particularly in classification of audio data. For better performance of audio classification, feature selection and combination play a key role as they have the potential to make or break the performance of any DL model. To investigate this role, we conduct an extensive evaluation of the performance of several cutting-edge DL models (i.e., Convolutional Neural Network, EfficientNet, MobileNet, Supper Vector Machine and Multi-Perceptron) with various state-of-the-art audio features (i.e., Mel Spectrogram, Mel Frequency Cepstral Coefficients, and Zero Crossing Rate) either independently or as a combination (i.e., through ensembling) on three different datasets (i.e., Free Spoken Digits Dataset, Audio Urdu Digits Dataset, and Audio Gujarati Digits Dataset). Overall, results suggest feature selection depends on both the dataset and the model. However, feature combinations should be restricted to the only features that already achieve good performances when used individually (i.e., mostly Mel Spectrogram, Mel Frequency Cepstral Coefficients). Such feature combination/ensembling enabled us to outperform the previous state-of-the-art results irrespective of our choice of DL model. 
\end{abstract}


\section{Introduction}

Audio data has been around us for a long time and is becoming an integral part of several cutting-edge computing and multimedia applications in several fields, e.g., security, healthcare monitoring, and context-aware services. The success of such applications stands on their capability to effectively store such data~\cite{ciritoglu2018towards} and perform audio related tasks such as classifying or retrieving audio files/signals (e.g., speech, music, environment sound/noise and other audio signals) based on their sound properties/content~\cite{gao2022multi}. 

While it has been, and it still is, a challenge for machines to accurately perform such audio related tasks, we are continuously devising better content-based classification and retrieval of audio databases to help machines perform these tasks~\cite{gao2022multi,xu2018large}--some of which are emerging as commercial products (e.g., findsounds.com and midomi.com) or part of larger applications (e.g., Google Hum to Search or voice recognition in virtual assistants).

Deep learning has been successful in audio classification~\cite{gao2022multi} with a tremendous amount of applications ranging from speech recognition~\cite{padmanabhan2015machine}, to music classification\cite{nanni2017combining}, and environmental sound classification~\cite{piczak2015environmental,piczak2015esc}. While earlier works have previously attempted to train Neural Networks using original audio data (e.g., raw audio signals, and standard low-level signal parameters)~\cite{lee2017sample}, more recent works have since observed that they could achieve significantly better performances by training the neural networks on extracted features that are tailored to the audio data at hand~\cite{palanisamy2020rethinking}. 

There have been many studies on audio content analysis, using
different features and different methods~\cite{palanisamy2020rethinking,schindler2018multi,li2019multi}. Despite the significant gains obtained by using extracted features, there is still a gap in terms of efficiency, reliability, and accuracy as most of existing methods use a single-modality along with the feature extraction.

Previous works demonstrated that the features fed to neural networks influence significantly the accuracy of the classification results. For instance, in the context of image classification, Wang et al.~\cite{wang2018locality} have shown that combining both spatial and spectral features improved greatly the classification accuracy. In our work, we seek to do the same for audio data, i.e., we would like to identify what combinations of features would enable different types of neural network models to achieve the best accuracy in audio classification. We particularly investigate the combination of three audio features (i.e., Mel Spectrogram~\cite{ball2006field}, Mel Frequency Cepstral Coefficients (MFCC~\cite{majeed2015mel}), and Zero Crossing Rate (ZCR~\cite{giannakopoulos2014introduction})), when used as ensembling with different deep learning models (i.e., Convolutional Neural Network (CNN), EfficientNet and MobileNet) on three benchmark speech classification datasets.

In this paper, we make the following contributions:
\begin{itemize}
    \item We explore different features and their ensembling for audio digit classification. 
    \item We investigate the best combination of features through a wide range of experiments using different models, on various datasets. 
    \item Our experiments suggest that our the proposed approach is effective in terms of both time and accuracy. 
    \item Finally, we release our source code and trained models for the research community to carry out the future research.
\end{itemize}

The rest of this paper is organised as follows. First, we present the context of our work, and in particular, we describe the related work and background of audio features (Section~2), then, we describe features ensembling approaches (Section~3), next, present the design of our experiments (Section~4), present evaluation (Section~5) and finally, we conclude this paper (Section~6).

\section{Background and Related Work}
\label{sec:background}
Audio classification has been a focus of a large number of works~\cite{7952132,sakashita2018acoustic,xu2018large,piczak2015esc} each leveraging different features including Mel Spectrogram (MS~\cite{ball2006field}), Mel Frequency Cepstral Coefficients (MFCC~\cite{majeed2015mel}) and Zero Crossing Rate (ZCR~\cite{giannakopoulos2014introduction}) or a combination of any two features as an ensemble. 

\subsection{Mel Spectrogram (MS)}
Audio signals are one dimensional, i.e., a time series of varying amplitudes. Since neural networks require fix dimensional inputs, it is necessary to convert/adapt audio signals into better formats which neural networks are able to process efficiently. One such format could be obtained by transforming audio signals into Mel Spectrogram~\cite{thornton2019audio,ball2006field} which have the advantage of providing the same information that the humans perceive. Spectrograms also provide a visual understanding of audio signals. Furthermore, Mel Scale is used to make the signal linear--matching with the human auditory system. Mathematically, it is formulated as in Equation~\ref{eq:mel_spectrogram}:

\begin{equation}
m=2595 \log _{10}\left(1+\frac{f}{700}\right)
\label{eq:mel_spectrogram}
\end{equation}
Where $m$ and $f$ represent Mel Spectrogram and frequency in Hz, respectively. 
Previously, many researchers have used Mel Spectrogram as feature for classification \cite{sakashita2018acoustic,mckinney2003features, park2019specaugment}. Sakashita and Aono~\cite{sakashita2018acoustic} compute Mel Spectrogram from different audio channels (i.e. Binaural, Mono, Harmonicpercussive source separation). Then, they segment the spectrogram into different flavours, and finally they train and ensemble many neural networks. McKinney and Breebaart~\cite{mckinney2003features} use features that incorporate Low-Level Signal Properties, Mel-Frequency Spectral Coefficients, and two other sets. Park et al.~\cite{park2019specaugment} apply three kinds of Log Mel Spectrograms including time wrapping, a deformation of the time series in the time direction, and the frequency masking. In their approach, the authors proposed a simple data augmentation method for speech recognition which is applied to listen, attend and spell networks for end-to-end speech recognition tasks. 

\subsection{Mel Frequency
Cepstral Coefficients (MFCC)}

The MFCC feature has been popular due to compressed representation of the signal~\cite{logan2000mel,majeed2015mel}. The computation of MFCC feature starts by segmenting audio signals into frames before taking discrete Fourier Transform and logs of
amplitude spectrum. Then, it performs Mel scaling and smoothing. Next, it takes a discrete cosine transform of the previous step to finally get the MFCC features. A detailed description of the features is provided by Logan~\cite{logan2000mel}. Like MS, many researchers used MFCC due to its compressed representation for audio classification (e.g.,~\cite{ittichaichareon2012speech, piczak2015esc,9616775}) in two ways: (i) extract MFCC then train different neural networks~\cite{ittichaichareon2012speech, piczak2015esc} or (ii) extract both MFCC and MS and use them to train two networks that are later ensembled~\cite{9616775}. 

\subsection{Zero Crossing Rate (ZCR)}
ZCR measures how signals change from positive to negative via zero or vice-versa~\cite{giannakopoulos2014introduction}. It helps to distinguish between highly correlated and uncorrelated features. Due to its correlation property, it is used by many researchers and it shows massive gain in performance~\cite{lu2000investigation,piczak2015esc}. 
Lu and Hankinson~\cite{lu2000investigation} use zero-crossing rate (ZCR) and its various combinations for the automatic audio indexing and retrieval systems. Whereas Piczak~\cite{piczak2015esc} provide a baseline performance using MFCC and ZCR as features. Both of these features drastically improved accuracy.

\subsection{Feature Ensembling}
Many works also tried ensemble features and networks~\cite{nanni2021ensemble,9616775,nanni2017combining,piczak2015environmental,zhang2017speech}. Nanni et al.~\cite{nanni2017combining} first get three features including spectrograms, a gammatonegram, and a rhythm  from audio input, and segment them into different windows, before training many SVM is trained and ensembling their predictions. Moreover, Nanni et al.~\cite{nanni2017combining} leverage multiple additional features and perform data augmentation to increase the data, then ensembled the multiple model predictions. Similarly, Niranjan et al.~\cite{9616775} use two extra features (i.e., MFCC and MS) for ensembling of CNN which showed a massive performance gain on the ESC50 dataset (i.e., a dataset for environmental sound classification). Some studies have attempted using multiple features to train their deep learning algorithms. Piczak~\cite{piczak2015environmental} devised a convolutional neural network for classifier training which combines two features (i.e., MFCC and its delta) whereas Zhang et al.~\cite{zhang2017speech} extract three Mel Spectrogram features (i.e., static, delta, and delta) for their training. 

So far, all the existing work in the literature with feature ensembling for audio data are only proposing and describing their approach with a unique configuration of features and model. Instead, in this paper, we explore different combinations of features with a diverse set of models, on different datasets. The goal of our work is to identify the best combination of features and models in terms of type and number.

\section{Ensembling Approach}
\label{sec:approach}
In this section, we explain our investigated ensembling approach (publicly available on \url{github.com/turab45/multi-features-ensembler-for-audio-classification}). The approach starts by taking input audio sample $X_{input}$. From $X_{input}$, we extract three different features namely, Mel Spectrogram \cite{ball2006field} $X_{mel_spe}$, Mel Frequency Cepstral Coefficients (MFCC~\cite{majeed2015mel})  $X_{MFCC}$, Zero Crossing Rate (ZCR~\cite{giannakopoulos2014introduction}) $X_{ZCR}$. After feature extraction, we propose the approach as described below:

\subsection{Multi-modality approach}
\begin{figure}[!htbp]
    \centering
    \includegraphics[width=1\textwidth]{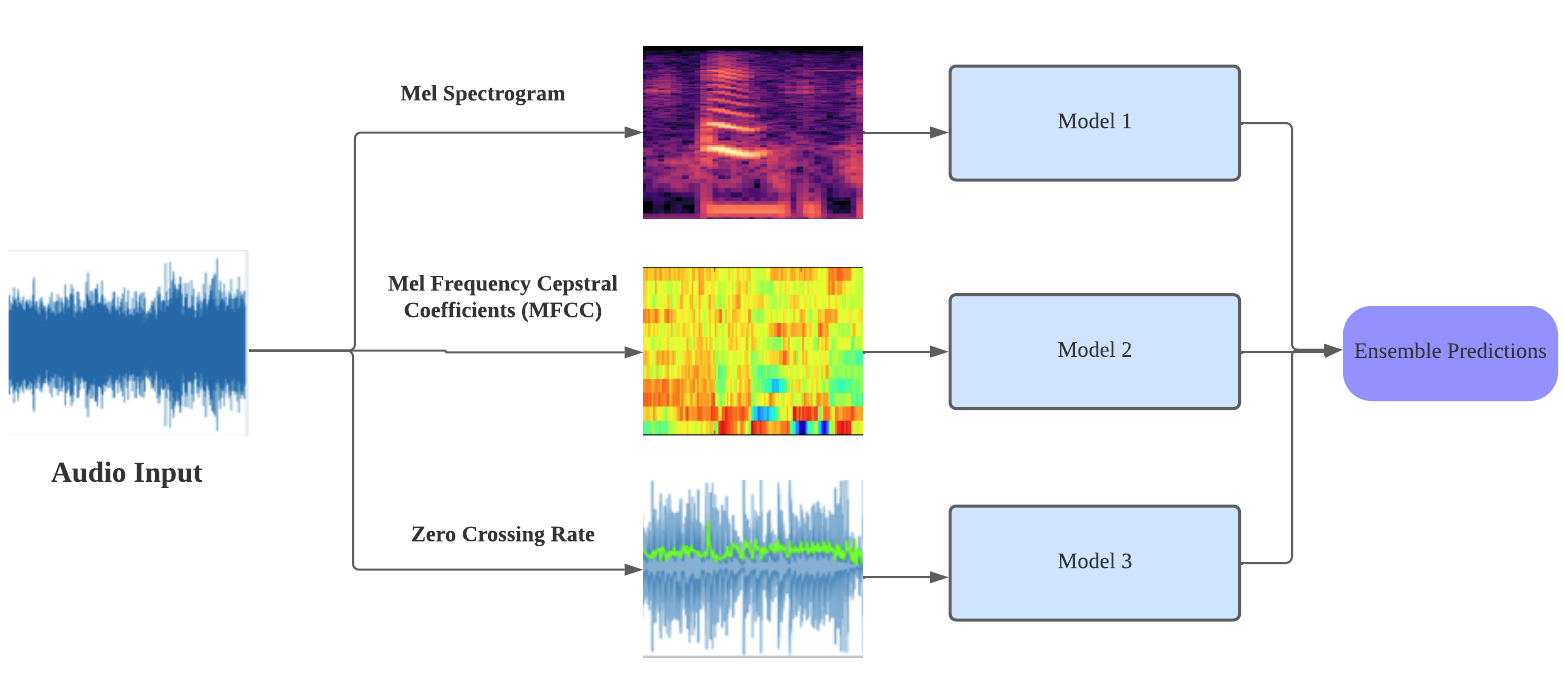}
    \caption{The investigated multi-modality approach with three models}
    \label{fig:multimodality_Arch}
\end{figure}

In this approach, we first train each model for each feature like $M_{MS}$, $M_{MFCC}$ and $M_{ZCR}$ for $X_{MS}$ $X_{MFCC}$ and $X_{ZCR}$, respectively. Once the models have been trained, we save those models. During test, sample $X_{test}$ is converted to three features and then passed to each model accordingly. Probability value of each model are average as ensembler. Then we predict the class for sample $X_{test}$, as shown in Figure~\ref{fig:multimodality_Arch}.

We describe the whole steps during test once models are trained:
\begin{itemize}
    \item Get probabilities:
    \begin{itemize}
        \item $P_{MS}$ = $M_{MS} (X_{MS})$
        \item $P_{MFCC}$ = $M_{MFCC} (X_{MFCC})$
        \item $P_{ZCR}$ = $M_{ZCR} (X_{ZCR})$
    \end{itemize}
    \item We then explore combination of two feature probabilities first, then we combined all these features probability, like 
    $Average=\frac{P_{MS} +P_{MFCC} +P_{ZCR}}{3}$
    \item Finally, we predict the class or the label using argmax function:
    \begin{itemize}
        \item $Predicted_{label} = argmax(Average)$ 
    \end{itemize}
\end{itemize}

\section{Experiment Design}
\label{sec:Experiment Design}
In this section, we describe our experimental design in three parts: (i) the data set on which we are basing our experiments, (ii) the algorithms we are comparing against, and (iii) the setup of our system and the values defined for the parameters of our algorithms.
\subsection{Datasets}
We used three different datasets for digit classification in English, Urdu and Gujrati languages. We describe each of the datasets below.
\subsubsection{Free Spoken Digits Dataset (FSDD~\cite{fssd})} This dataset is about spoken digits of English pronunciation (0 to 9)  and consists of 1500 recordings.  These are recorded from 3 different speakers and 50 of each digit per speaker. Each recording is mono and sampled at  8kHz and saved in \emph{.wav} format. It has 10 classes (0 to 9). All participants are male. Information about audio duration is unknown, but we noticed that its minimum duration length is 1 second and maximum is 2 second.

\subsubsection{Audio Urdu Digits Dataset (AUDD~\cite{aiman2021audd})} 
This dataset is an audio spoken digits dataset for Urdu language. This dataset has 25218 samples collected from 740 participants aged between 5 an 89 for diversity purpose, but the majority of participants were 5 to 14 years old and male participants were slightly more numerous than female participants. Each sample is stereo and sampled at 48 $kHz$ and is mono-channel with a minimum length of 1 second and maximum length of 2 seconds. Furthermore, the number of samples in the dataset is almost balanced between all the classes (i.e., digit per age).

\subsubsection{Audio Gujarati Digits Dataset (AGDD~\cite{dalsaniya2019development})} 
This dataset is an audio digits dataset for Gujarati language which has 1940 samples sampled at 44.1 kHz. Recordings are obtained by 20 users, including 14 male and 6 female, from five different regions of Gujarat i.e. Central Zone, North Zone, South Zone, Saurashtra, Kutch Region. Information about duration of audio is unknown but we notice that minimum and maximum duration of audio is 1 and 2 seconds, respectively. Each audio is saved in .wav format.

\subsection{Algorithms}
We used different deep learning models for generalization purpose. We used 3-layers CNN, EfficientNet~\cite{tan2019efficientnet} and mobileNet~\cite{howard2017mobilenets}. For a fair comparison, we took the same model architectures that were defined in~\cite{aiman2021audd,dalsaniya2019development}. For the CNN model, we implemented a 3-layer architecture as described in ~\cite{aiman2021audd} and shown in Table~\ref{architecture}.

\begin{table}[htp]\small 
\centering
\caption{CNN Model, architecture taken from ~\cite{aiman2021audd}}
\label{architecture}
\begin{tabular}{|l|l|l|}
\hline \textbf{Layer type} & \textbf{Dimensions} & \textbf{Other Details} \\
\hline  Input & Input layer & $(32,32,1)$  \\
\hline CNN & $(30,30,64)$ & kernel $3 \times 3 ;$ stride 1 ; \\ & & relu  activation  \\
\hline Max Pool & $(15,15,64)$ &  N/A \\ 
\hline BN & $(15,15,64)$ & default value as \\ & & given in Keras \cite{chollet2015keras} \\

\hline CNN & $(13,13,64)$ & kernel $3 \times 3;$ stride 1; \\ & & relu activation  \\
\hline Max Pool & $(6,6,64)$ &  N/A \\ 
\hline BN & $(6,6,64)$ & default value as \\ & & given in Keras \cite{chollet2015keras} \\

\hline CNN & $(4,4,64)$ & kernel $3 \times 3 ;$ stride 1; \\ & &  relu activation  \\
\hline Max Pool & $(2,2,64)$ & N/A \\ 
\hline BN & $(2,2,64)$ & default value as\\ & & given in Keras \cite{chollet2015keras} \\

\hline Dropout & $(2,2,64)$ & dropout rate=0.1 \\
\hline Flatten & 256 & N/A \\
\hline Fully Connected & 512 & N/A \\
\hline Dropout & 512 & dropout rate=0.1 \\

\hline Fully Connected & 128 & N.A \\
\hline Dropout & 512 & dropout rate=0.1 \\

\hline Fully Connected & 10 &  softmax activation\\
\hline

\end{tabular}

\end{table}

\subsection{Setup}
Once features are extracted by using MS, MFCC, ZCR, we reshape the features to $32 \times 32 \times 1$ as 2D images. Each dataset is randomly split into training and test sets with an 8 to 2 ratio. Furthermore, the training set is also split into training and validation sets with a 9 to 1 ratio. 
 
We used hyperparameters as epoch 150, learning rate 0.01, batch size of 64 and set the loss to categorical cross entropy as shown in Equation~\ref{eqn:CrossEntropy}. 
 \begin{equation}
\label{eqn:CrossEntropy}
L(\Theta)=-\sum_{i=1}^{k} y_{i} \log \left(\hat{y}_{i}\right) 
\end{equation}
where $y_{i}$ is the ground truth,  $\hat{y_i}$ is the predicted label,  $k$ is the number of samples in the batch.
\section{Evaluation} \label{sec:evaluation}
We evaluated the performance using accuracy metric (as shown in Equation~\ref{eqn:accuracy}) and time required for testing (in millisecond).

\begin{equation}
\label{eqn:accuracy}
A=\frac{P}{T} 
\end{equation}
where $A$, $P$ and $T$ represent accuracy, correct number of predicted samples and total number of samples, respectively.  \\ 
We performed each experiment three times and average accuracy is reported. In Tables~\ref{Audd_Accuracy}, \ref{fssd_accuracy} and \ref{Gujrati_accuracy}, the values MS, MFCC and ZCR represent Mel Spectrogram,
Mel Frequency Cepstral Coefficients, and
Zero Crossing Rate, respectively. In braces, MS means the experiment is performed using a single feature and this is the same for MFCC and ZCR. For two or three features in braces, the model prediction is obtained using an ensemble of models trained using two or three features. EfficientNetB0 to EfficientNetB7 are the different versions of EfficientNet and similarly we used a single version of MobileNetV1. 

Tables~\ref{Audd_Accuracy}, \ref{fssd_accuracy} and \ref{Gujrati_accuracy}, also report the average time taken for testing (in milliseconds). We evaluated the testing time for each feature individually and when combined as an ensemble, with each of the considered models and datasets. 

Obtained results suggest that testing time is dependent of all dataset, model and feature. For the same dataset, time for CNN with MS is lower than with ZCR. For the same model, time of CNN with MFCC is larger than CNN with MS. For the same feature, time of CNN with MS on AUDD is lower than CNN with MS on FSSD. Otherwise, it is natural that when increasing the number of features, time increases as well. So time for two and three features increases. Overall, obtained results suggest that if we want to achieve a trade-off between accuracy and speed, we should also select our features and models carefully.

For AUDD dataset~\cite{aiman2021audd}, among single feature experiments, CNN with MS feature achieves the highest performance, whereas the same CNN with the ZCR feature achieves the worst performance. That is due to the loss of sound features during ZCR extraction. 

While we are not able to identify a rule on what number of features is ideal during emsembling (i.e., either two or three features at a time), we see that the features showing high performance independently lead to better performance when combined in a feature ensembling. ZCR which achieved less performance acts as a handicap as it drops the performance whenever it is combined with MFCC or MS. MFCC and MS combination with CNN model showed the best performance over previous SOTA performance with an absolute improvement of 3.00\%, as shown in Table~\ref{Audd_Accuracy}.

Similarly for the FSSD dataset, we performed many experiments using diverse DL models. First we explore single feature based performances. Then, we check combinations of feature ensemblings. Unlike ZCR for AUDD, ZCR for FSSD a shows better performance. Therefore, the combination ZCR with MFCC and MS using either CNN or EfficientNet helps to improve the performance during ensembling and shows superior performance with a 1\% improvement to the previous SOTA performance, as shown in Table~\ref{fssd_accuracy}. 

Moreover, to check the effectiveness of the approach, we perform the experiment on Audio Gujarati Digits Dataset (AGDD). In AGDD dataset, MS single feature has been effective with CNN and similarly MFCC is also effective. When these two features are used during ensembling, it further improved the performance over SOTA performance with an absolute improvement of 0.2\%. However, these two individual features with EfficientNet and MobileNet  have shown worse performance than CNN. None of the different DL models using three features ensembling were unable to improve performance due to the presence of the ZCR feature which acted as a handicap--achieving worse performance compared to MS and MFCC features. 

Overall, MS and MFCC combination showed the best performance for audio digit classificaiton, nevertheless model selection is very important. MS and MFCC combination with EfficientNet and MobileNet unable to improve the performance. To select the good features, we should also consider the model choice. ZCR with MS and MFCC ensembling is only good choice of selection, whenever it shows good performance individually as shown in Table~\ref{Audd_Accuracy}. In the most cases, MS and MFCC has shown the best performance and results suggest that we should use those two features combination for ensembling.

 \begin{table}[htp!]
\centering
\caption{Different features accuracy using different models using AUDD~\cite{aiman2021audd} dataset}
\label{Audd_Accuracy}
\begin{tabular}{|l|c|c|}
\hline
\textbf{Model Name} & \multicolumn{1}{l|}{\textbf{Accuracy}} & \multicolumn{1}{l|}{\textbf{ Time(ms)}} 
\\ \hline
\multicolumn{3}{|c| }{\textbf{ Single feature  }}
\\ 
\hline

Support Vector Machine (MS) \cite{aiman2021audd}       &   0.65 $\pm$ 0.0 & --
                                     \\ \hline
Multilayer Perceptron (MS) \cite{aiman2021audd}      & 0.73 $\pm$ 0.02  & --                                       
\\ \hline  CNN (MS) \cite{aiman2021audd}  & 0.86 $\pm$ 0.02 & 0.264
\\ \hline EfficientNetB0  (MS) \cite{aiman2021audd}  &0.84$\pm$ 0.05& 1.60
\\ \hline EfficientNetB1 (MS) \cite{aiman2021audd} &0.82$\pm$ 0.02& --
\\ \hline EfficientNetB2 (MS) \cite{aiman2021audd} &0.83$\pm$ 0.04& --
\\ \hline EfficientNetB3 (MS)  \cite{aiman2021audd} &0.84$\pm$ 0.06& --
\\ \hline EfficientNetB4 (MS) \cite{aiman2021audd} &0.82$\pm$ 0.03& --
\\ \hline EfficientNetB5 (MS)\cite{aiman2021audd}  &0.84$\pm$ 0.04& --
\\ \hline EfficientNetB6 (MS) \cite{aiman2021audd} &0.81$\pm$ 0.06& --
\\ \hline EfficientNetB7 (MS) \cite{aiman2021audd} & 0.56$\pm$ 0.07& --
\\ \hline MobileNetV1 (MS) &  0.83$\pm$ 0.00 & 0.531

\\ \hline  CNN (MFCC)  & 0.85 $\pm$ 0.01 & 0.541
\\ \hline  CNN (ZCR)  & 0.40 $\pm$ 0.03 & 0.360
\\ \hline EfficientNetB0 (MFCC)  &0.81$\pm$ 0.03 & 1.124
\\ \hline EfficientNetB0 (ZCR)  &0.27 $\pm$ 0.02 & 1.681
\\ \hline MobileNetV1 (MFCC)  &0.80$\pm$ 0.02 & 0.628
\\ \hline MobileNetV1 (ZCR)  &0.36 $\pm$ 0.03 & 0.689

\\ \hline
\multicolumn{3}{|c| }{\textbf{ Two Features Ensembler }}
\\ \hline  CNN (MS and MFCC)  & 0.89 $\pm$ 0.01 & 0.896
\\ \hline  CNN (MS and ZCR)  & 0.82 $\pm$ 0.03 & 0.805
\\ \hline  CNN (MFCC and ZCR)  & 0.81 $\pm$ 0.01 & 0.631

\\ \hline EfficientNetB0 (MS and MFCC )  &0.88$\pm$ 0.02  & 4.431
\\ \hline EfficientNetB0 (MS and ZCR)  &0.82$\pm$ 0.03  & 3.899
\\ \hline EfficientNetB0 (MFCC and ZCR)  &0.80$\pm$ 0.02  & 4.076

\\ \hline MobileNetV1 (MS and MFCC )  &0.87$\pm$ 0.03   & 1.65
\\ \hline MobileNetV1 (MS and ZCR)  &0.66$\pm$ 0.02   & 1.70
\\ \hline MobileNetV1 (MFCC and ZCR)  &0.72$\pm$ 0.00   & 1.13

\\ \hline
\multicolumn{3}{|c| }{\textbf{ Three Features Ensembler }}
\\ \hline  CNN (MS, MFCC and ZCR)  & 0.87 $\pm$ 0.03   & 0.8167
\\ \hline EfficientNetB0 (MS, MFCC and ZCR)  &0.86$\pm$ 0.03    & 5.859
\\ \hline MobileNetV1 (MS, MFCC and ZCR)  &0.85$\pm$ 0.05   & 2.540

\\ \hline
\end{tabular}

\end{table}

\begin{table}[htp!]
\centering
\caption{Performance comparison multiple features using different models on FSSD \cite{fssd} }
\label{fssd_accuracy}
\begin{tabular}{|l|c|c|}
\hline
\textbf{Model Name} & \multicolumn{1}{l|}{\textbf{Accuracy}} & \multicolumn{1}{l|}{\textbf{Time(ms)}}
\\ 
\hline 

CNNDigitReco-speakerindependent~\cite{oscar2020} & 0.78 $\pm$ 0.0 & --
\\ \hline
\multicolumn{3}{|c| }{\textbf{ Single Feature }}
\\ 
\hline 
Support Vector Machine~\cite{inamulhaq2021} &  0.90 $\pm$ 0.02 & --
\\
\hline
 Random Forest \cite{inamulhaq2021} & 0.96 $\pm$ 0.06 & --

\\  \hline  English Digit Model\cite{nasr2021text}  & 0.97 $\pm$ 0.09 & -- 
\\ \hline CNN (MS) \cite{aiman2021audd} & 0.973 $\pm$ 0.01 & 0.3441  
\\ \hline CNN (MFCC)  & 0.978 $\pm$ 0.03 &   0.2895
\\ \hline CNN (ZCR)  & 0.572 $\pm$ 0.08 &   0.306

\\ \hline EfficientNetB0 (MS) & 0.947 $\pm$ 0.05 &   1.154
\\ \hline EfficientNetB0 (MFCC)  & 0.968 $\pm$ 0.07 &  0.740 
\\ \hline EfficientNetB0 (ZCR)  & 0.378 $\pm$ 0.06 &   1.148

\\ \hline MobileNetV1 (MS) & 0.877 $\pm$ 0.01  &   4.855
\\ \hline MobileNetV1 (MFCC)  & 0.980 $\pm$ 0.02 & 3.367    
\\ \hline MobileNetV1 (ZCR)  & 0.538 $\pm$ 0.03 &  4.035 

\\ \hline
\multicolumn{3}{|c| }{\textbf{ Two Feature Ensembler }}
\\ \hline CNN (MS and MFCC)  & 0.987 $\pm$ 0.02 & 0.651
\\ \hline CNN (MS and ZCR)  & 0.980 $\pm$ 0.03  & 0.630
\\ \hline CNN (MFCC and ZCR)  & 0.977 $\pm$ 0.05  & 0.639

\\ \hline EfficientNetB0 (MS and MFCC)  & 0.987  $\pm$ 0.07 & 2.811
\\ \hline EfficientNetB0 (MS and ZCR)  & 0.957 $\pm$ 0.08  & 3.583
\\ \hline EfficientNetB0 (MFCC and ZCR)  & 0.970  $\pm$ 0.01 & 2.838

\\ \hline MobileNetV1 (MS and MFCC)  & 0.985 $\pm$ 0.05  & 4.9 
\\ \hline MobileNetV1 (MS and ZCR)  & 0.960 $\pm$ 0.06  & 1.682
\\ \hline MobileNetV1 (MFCC and ZCR)  & 0.970 $\pm$ 0.09  & 2.308
\\ \hline

\multicolumn{3}{|c| }{\textbf{ Three Features Ensembler}}

\\ \hline CNN (MS, MFCC, ZCR)  & 0.99 $\pm$ 0.03 & 0.921
\\ \hline EfficientNetB0 (MS, MFCC, ZCR)  & 0.99 $\pm$ 0.04 & 4.977
\\ \hline MobileNetV1 (MS, MFCC, ZCR)  & 0.987 $\pm$ 0.05 & 2.903

\\ \hline
\end{tabular}

\end{table}

\begin{table}[htp!]
\centering
\caption{Performance comparison multiple features using different models on Gujarati Digits dataset\cite{dalsaniya2019development} }
\label{Gujrati_accuracy}
\begin{tabular}{|l|c|c|}
\hline
\textbf{Model Name} & \multicolumn{1}{l|}{\textbf{Accuracy}} & \multicolumn{1}{l|}{\textbf{Time(ms)}}
\\ \hline
\multicolumn{3}{|c| }{\textbf{ Single Feature}}

\\ \hline  Gujarati Digits Model\cite{dalsaniya2019development}  & 0.75 $\pm$ 0.02 & -- 
\\ \hline CNN (MS) \cite{aiman2021audd} & 0.970 $\pm$ 0.03 & 1.770
\\ \hline CNN (MFCC)  &0.959 $\pm$ 0.05 &  1.761
\\ \hline CNN (ZCR)  &0.572 $\pm$ 0.07  &  1.750

\\ \hline EfficientNetB0 (MS)  &0.880 $\pm$ 0.02  &  9.402
\\ \hline EfficientNetB0 (MFCC)  &0.907 $\pm$ 0.01  &  9.781
\\ \hline EfficientNetB0 (ZCR)  &0.321 $\pm$ 0.04  &  9.536

\\ \hline MobileNetV1 (MS)   &0.856 $\pm$ 0.01   &  0.63
\\ \hline MobileNetV1 (MFCC)  &0.89 $\pm$ 0.05  & 0.791
\\ \hline MobileNetV1 (ZCR)  &0.557 $\pm$ 0.03  & 1.35

\\ \hline
\multicolumn{3}{|c| }{\textbf{ Two Features Ensembler }}
\\ \hline CNN (MS and MFCC)  &0.972 $\pm$ 0.04  & 2.996 
\\ \hline CNN (MS and ZCR)  &0.936 $\pm$ 0.08  &  1.006
\\ \hline CNN (MFCC and ZCR)  &0.943 $\pm$ 0.07  & 1.723

\\ \hline EfficientNetB0 (MS and MFCC)  &0.970 $\pm$ 0.01  & 3.491
\\ \hline EfficientNetB0 (MS and ZCR)  &0.881 $\pm$ 0.06  &  3.338
\\ \hline EfficientNetB0 (MFCC and ZCR)  &0.94 $\pm$ 0.01  & 3.334

\\ \hline MobileNetV1 (MS and MFCC)  & 0.933 $\pm$ 0.03  & 1.84
\\ \hline MobileNetV1 (MS and ZCR)  & 0.80 $\pm$ 0.04  & 2.396
\\ \hline MobileNetV1 (MFCC and ZCR)  &0.884 $\pm$ 0.02  & 2.280

\\ \hline
\multicolumn{3}{|c| }{\textbf{ Three Features Ensembler }}
\\ \hline CNN (MS, MFCC and ZCR)  &0.966 $\pm$ 0.03  & 1.254
\\ \hline EfficientNetB0 (MS, MFCC and ZCR)  &0.948 $\pm$ 0.05  &  5.261
\\ \hline MobileNetV1 (MS, MFCC and ZCR)  &0.917 $\pm$ 0.03  & 3.520

\\ \hline
\end{tabular}

\end{table}

\begin{figure}[!htbp]
\begin{tabular}{c}
     \includegraphics[width=0.75\linewidth]{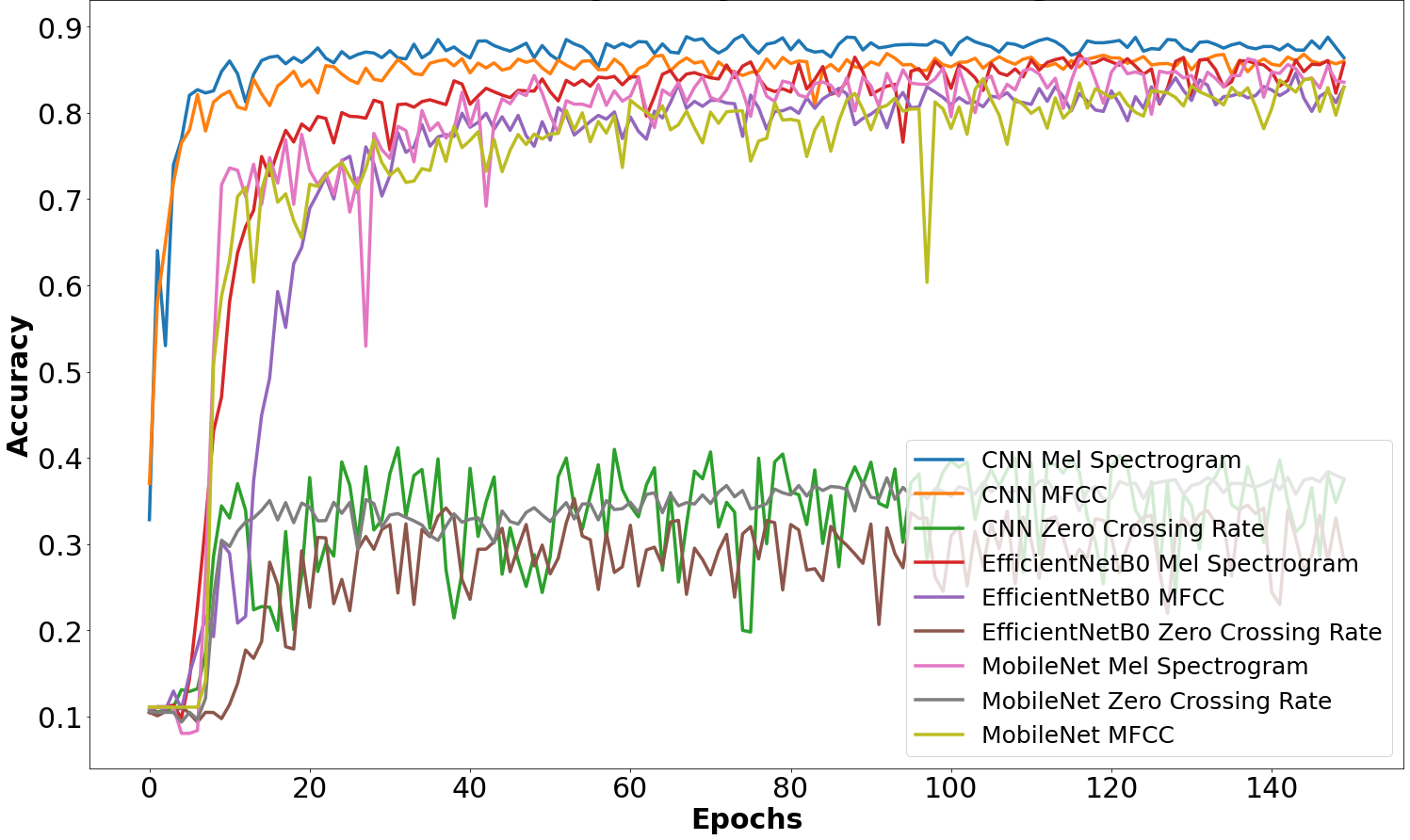}\\
     (a) Urdu digits Dataset \\
     \includegraphics[width=0.75\linewidth]{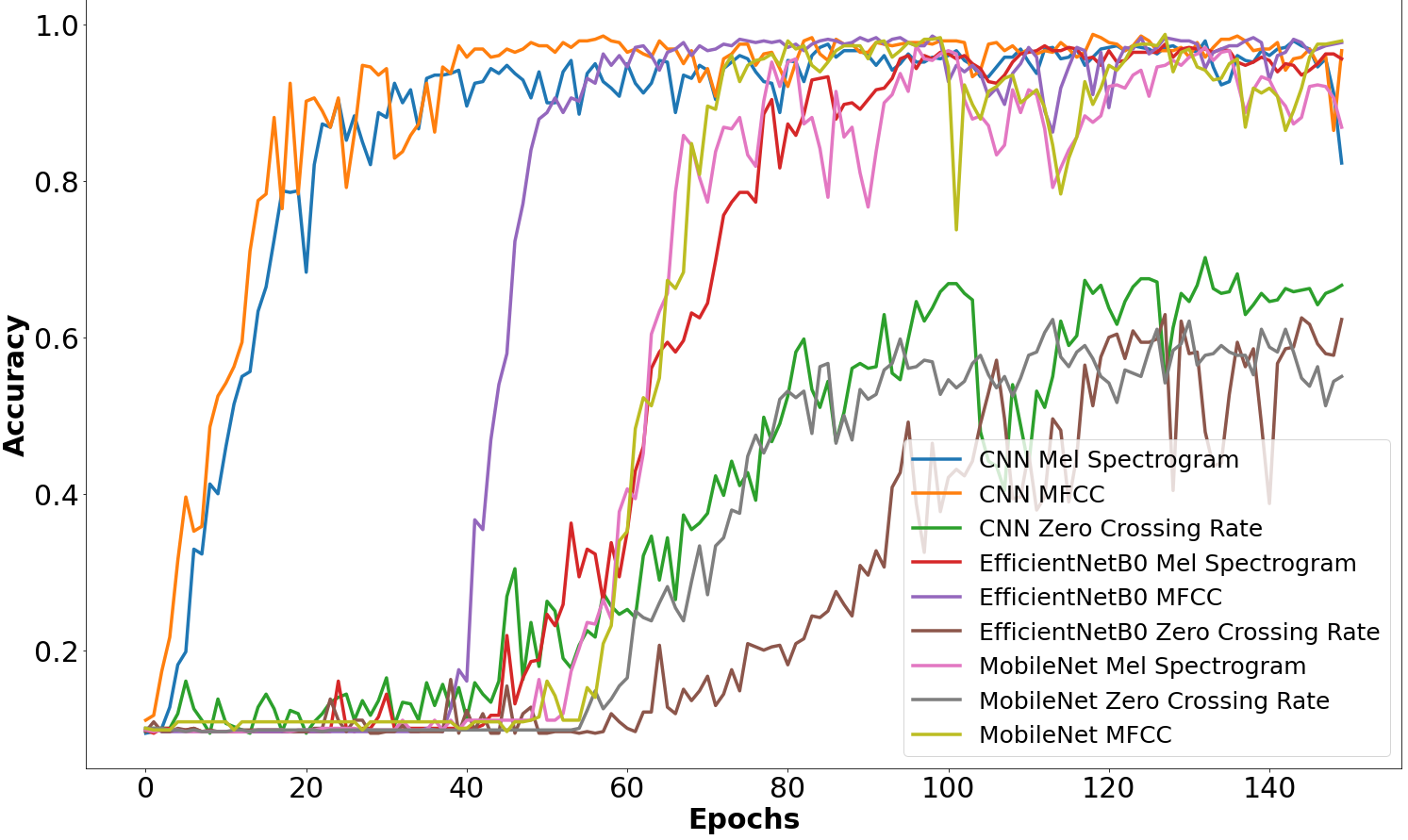}\\
     (b) FSDD Dataset  \\
     \includegraphics[width=0.75\linewidth]{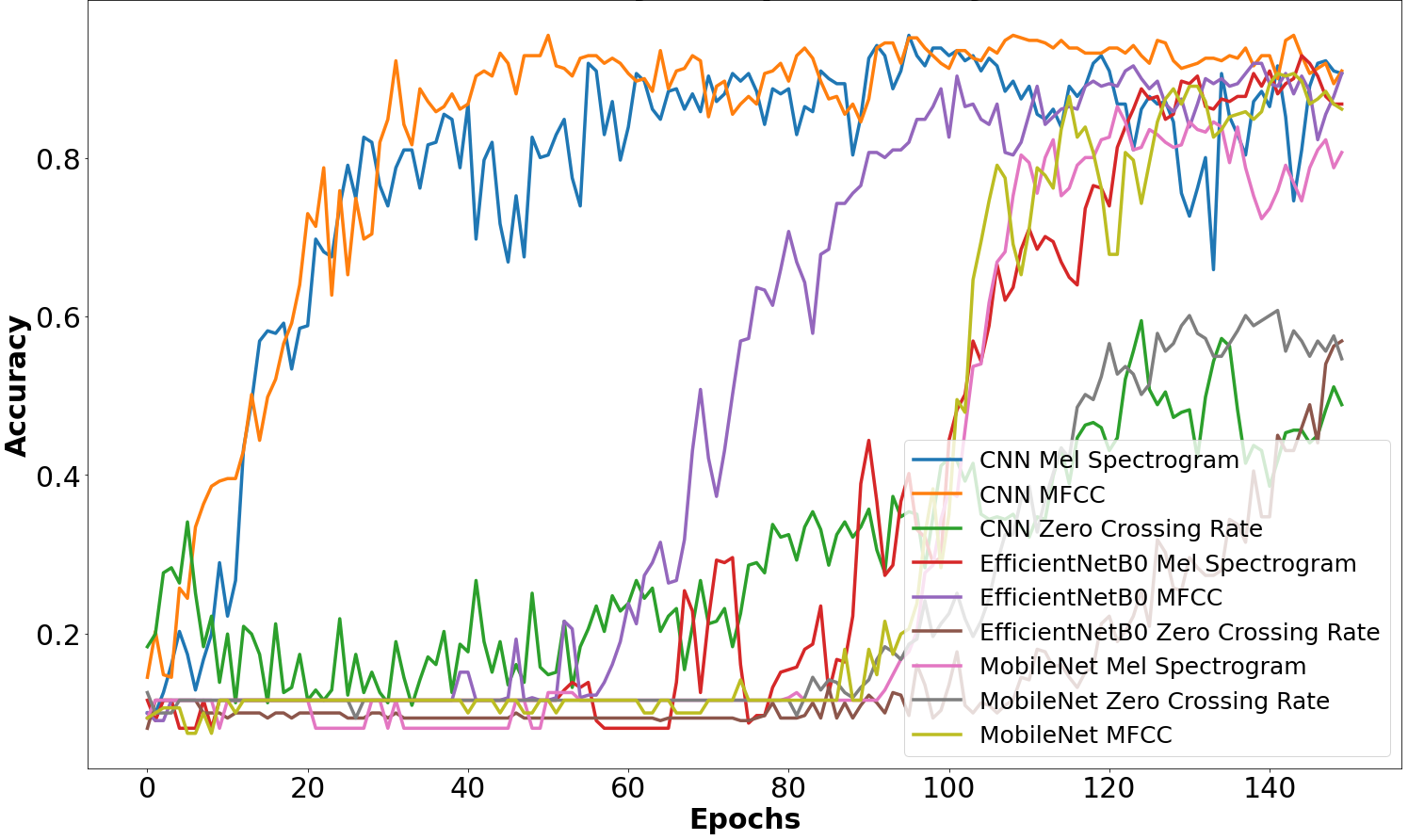}\\
     (c) Gujarati Dataset
\end{tabular}
    \centering
    
    \caption{Validation Accuracy over Epochs on the Different Datasets}
    \label{fig:validationoverepochs}
\end{figure}

We have also analyzed the stability of the models with a single feature at a time and their behavior during validation on AUDD, FSSD, and AGDD respectively, as shown in Figures~\ref{fig:validationoverepochs}.

Interestingly, all Figures~\ref{fig:validationoverepochs} have the same pattern as CNN with the MS feature, i.e., they are very stable. Furthermore, the analysis shows consistent accuracy improvements from one epoch to another, except for EfficientNet with ZCR which has shown a worse performance. Feature-wise, MS and MFCC are nearly equally successful among all cases with the different models, whereas the use of ZCR drops the performance of all models. 

Overall, Figures~\ref{fig:validationoverepochs} suggest that to achieve the best performance and stability, the choice of model is important, whereas MS seems to be an excellent feature extractor and the feature of choice. 

In summary, we have shown that ensembling all features at the same time does not guarantee achieving the best performance (in the contrary, it acts as a handicap) and that beyond the selection of features, the choice of the model is important. Therefore, it will be beneficial to design automatic and interpretable ensembling techniques, potentially through reinforcement learning techniques such as grammar-guided genetic programming~\cite{lynch2019evolutionary,saber2020evolving,saber2020wcci_scheduling,saber2018tpnc,saber2018genp}. Furthermore, while we have only focused on accuracy in this work, it is possible that a feature/model ensembling does not achieve the best accuracy, but performs better on other metrics. Therefore, it will be beneficial to formulate the problem as a multi-objective feature selection~\cite{saber2021reparation,saber2021incorporating,saber2018seeding,saber2018vm}.

\section{Conclusion} 
\label{sec:conclusion}
This paper investigates the use of multi-feature ensemblers, combining three state-of-the-art audio features (i.e., Mel Spectrogram, Mel Frequency Cepstral Coefficients, and Zero Crossing Rate) to alleviate the performance constrained by the features when dealing with audio classification tasks. 

In our work, we sought to explore different combinations of the three state-of-the-arts features with a diverse set of models, on different datasets with the goal of identifying the best combination of features and models. To check the generalization of our results, we used three different audio datasets including, Free Spoken Digits Dataset, Audio Urdu Digits Dataset and Audio Gujarati Digits Dataset.  

We trained our models with each feature individually, then with a combination of two and three features. We evaluated the performance of each configuration, i.e., model and feature(s) in terms of accuracy and testing time. Our thorough experimental evaluation has shown that it is only better to combine features that already perform well individually (i.e., mostly Mel Spectrogram, Mel Frequency Cepstral Coefficients). 

Our future research direction is in two folds: (i) to reduce testing time by stacking multiple features and (ii) to explore these features from data augmentation perspective to generate novel features.

\section{Acknowledgment}
This research was supported by Science Foundation Ireland (SFI) under grant numbers 18/CRT/6223, 13/RC/2094$\_$P2 (Lero SFI Centre for Software) and 13/RC/2106$\_$P2 (ADAPT SFI Research Centre for AI-Driven Digital Content Technology).

\bibliographystyle{plain}
\bibliography{references.bib}

\section*{Authors}
\vspace{1cm}

\begin{tabular}{p{3cm} p{10cm}}
    \includegraphics[width=25mm,keepaspectratio,valign=T]{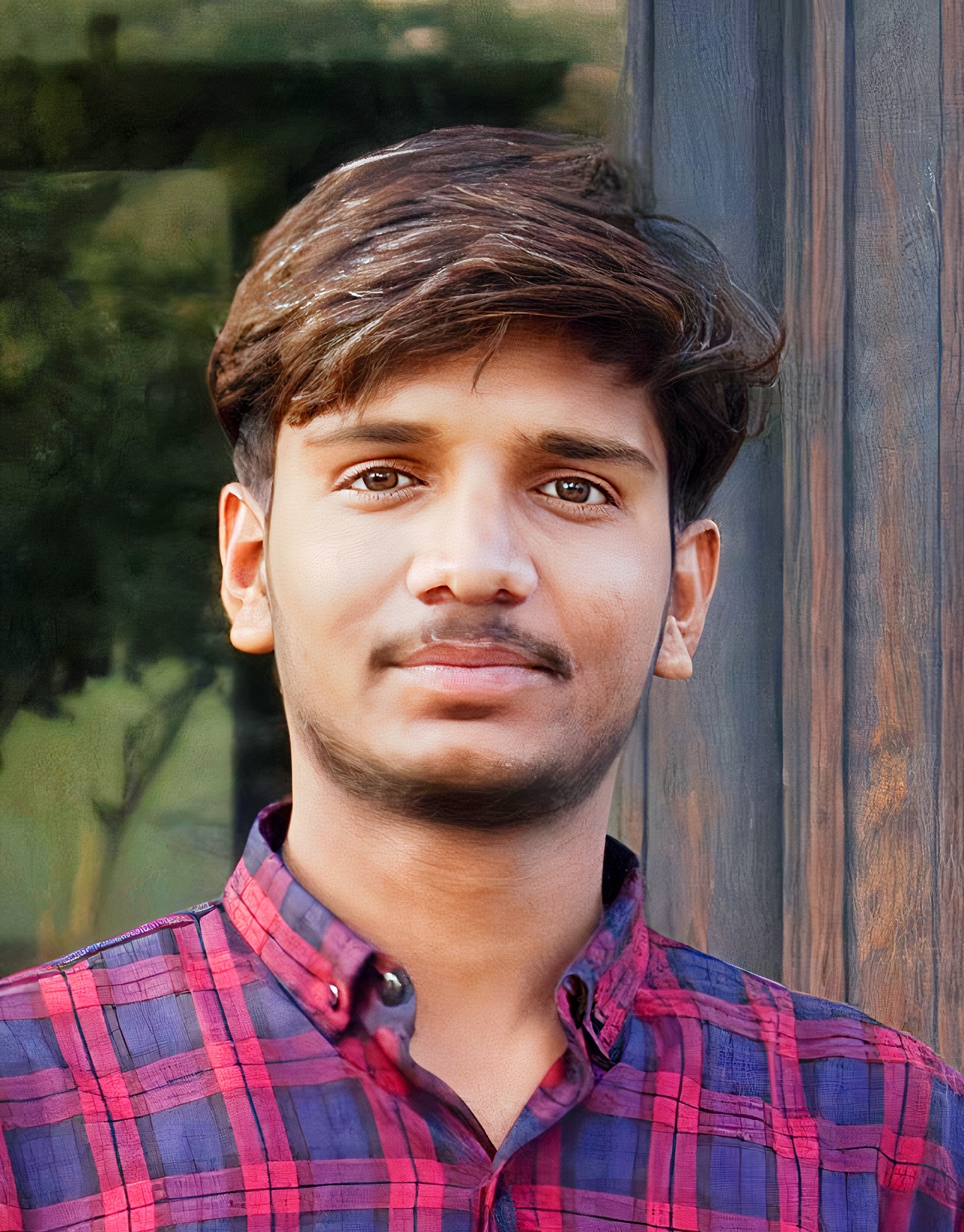} & \textbf{Muhammad Turab} is an undergraduate student at MUET, Jamshoro. He has been working in the field of Deep Learning and Machine learning for two years. He has completed more than 70 projects on GitHub. His research interests include deep learning, computer vision and data augmentation for medical imaging.
    \\ \\
    
    \includegraphics[width=25mm,keepaspectratio,valign=T]{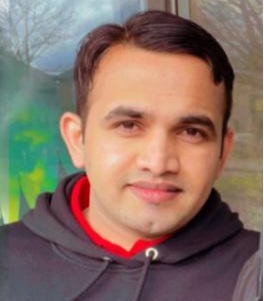} & \textbf{Teerath Kumar} received his Bachelor’s degree in Computer Science with distinction from National University of Computer and Emerging Science (NUCES), Islamabad, Pakistan, in 2018. Currently, he is pursuing PhD from Dublin City University, Ireland. His research interests include advanced data augmentation, deep learning for medical imaging, generative adversarial networks and semi-supervised learning.
    \\ \\
    
    \includegraphics[width=25mm,keepaspectratio,valign=T]{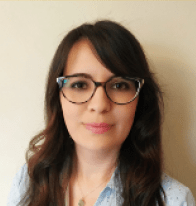} & \textbf{Malika Bendechache} is a Lecturer/Assistant Professor in the School of Computing at Dublin City University, Ireland, and a Funded Investigator at both ADAPT and Lero Science Foundation Ireland research centres. Malika holds a PhD in Computer Science from University College Dublin, Ireland. Her research interests span the areas of Big data Analytics, Machine Learning, Data Governance, Parallel and Distributed Systems, Cloud/Edge/Fog Computing, Blockchain, Security, and Privacy.
    \\ \\
    \includegraphics[width=25mm,keepaspectratio,valign=T]{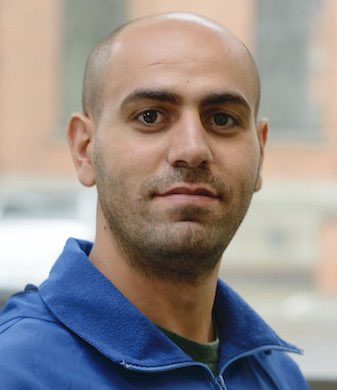} & \textbf{Takfarinas Saber} is a Lecturer/Assistant Professor in the School of Computer Science at National University of Ireland, Galway, Ireland, and a Funded-Investigator in Lero, the Science Foundation Ireland Research Centre for Software. Takfarinas holds a PhD in Computer Science from University College Dublin, Ireland. His area of expertise is in the optimisation of Complex Software Systems such as Cloud Computing, Smart Cities, Distributed Systems, and Wireless Communication Networks.
\end{tabular}

\end{document}